\documentstyle[epsfig]{jfm} 

\title[Very viscous drops cannot break up]
{Very viscous drops cannot break up}

\author[J. Eggers and M.A. Fontelos]
{Jens Eggers$^*$ and Marco A. Fontelos$^\dagger$
  }
\affiliation{
$^*$School of Mathematics, 
University of Bristol, University Walk, \\
Bristol BS8 1TW, United Kingdom 
$^\dagger$ Departamento de Matem\'atica Aplicada,
Universidad Rey Juan Carlos,
C/ Tulip\'an S/N, 28933 M\'ostoles, Madrid, Spain. \\
           }

\begin{document}

\maketitle

\begin{abstract}
We consider an axisymmetric, freely suspended fluid drop with 
surface tension, whose viscosity is so large that both inertia and forcing 
by an external fluid can be ignored. We show that whatever be the 
initial condition, pinchoff can never occur. 
\end{abstract}

\begin{figure}
\begin{center}
\psfig{figure=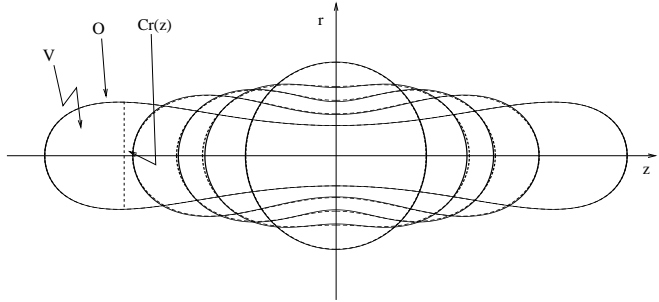,height=4cm}
\end{center}
\caption{
A viscous drop of unperturbed radius $R$ is initially extended to a 
length of $15.5\;R$. The full lines show it retracting back into
a sphere following Stokes' equation, at 
$t = n\times 0.646\; \eta R/\gamma, n=0,1,2,3$.
The dashed lines are profiles computed on the 
basis of the  lubrication equation (\ref{lub}), shown at
the same times. Note that the minimum local drop radius is 
always a monotonically increasing function.
       }
\label{fig1}
\end{figure}

The breakup of fluid drops has been studied very extensively 
(\cite{S94,E97,B02}) owing to its relevance to mixing, printing, 
and DNA analysis. In most circumstances, breakup of an extended drop
occurs almost inevitably owing to the Rayleigh instability 
(\cite{E97}), which tends to locally reduce the radius until breakup 
occurs. So one might think that a sufficiently extended drop, 
that has separated from a nozzle or has been stretched by an 
external fluid of comparatively low viscosity, will break up in
the same manner. Here we show that this is not the case, but that
the drop always retracts to its spherical state of minimum surface
energy, if inertial and external forces can be neglected, see 
figure \ref{fig1}. \cite{S94}, p.81, makes some prescient remarks 
which amount to the same statement, but we are not aware that
the impossibility of breakup has ever been shown.

The interior of the drop is thus described by Stokes' equation,
subject to a normal stress $\gamma\kappa {\bf n}$,
where $\gamma$ is the coefficient of surface tension 
and $\kappa=1/h(1+h_z^2)^{1/2}-h_{zz}/(1+h_z^2)^{3/2}$ 
twice the mean curvature of the interface. If $\sigma$ 
is the stress tensor, this can be summarised concisely by 
\begin{equation}
{\bf \nabla\cdot\sigma}=0\quad\mbox{in the drop}, \quad
{\bf \sigma n}=-\gamma{\bf n}\kappa\quad\mbox{on the surface}.
\label{Stokes}
\end{equation}
Integrating ${\bf \nabla\cdot
\sigma}$ over a volume $V$ bounded by the drop
surface and a plane perpendicular to the axis (cf. figure \ref{fig1}),
we find from the divergence theorem and from the boundary condition 
that 
\begin{equation}
0=\int_S{\bf n\sigma} ds= \int_{Cr(z)}{\bf n}\sigma ds
+ \int_{O}{\bf n\sigma} ds = \int_{Cr(z)}{\bf e}_z\sigma ds -
\gamma\int_{O}{\bf n}\kappa ds,
\label{slice}
\end{equation}
where $O$ is the surface as shown in figure \ref{fig1}, and 
$Cr(z)$ is the cross section of the drop at $z$.

Using ${\bf n}=(-h_z{\bf e}_z+{\bf e}_r)/(1+h_z^2)^{1/2}$, the 
integral over $O$ can be evaluated as
\begin{equation}
-2\pi\gamma{\bf e}_z\int_{z_{end}}^z
h h_z \kappa dz = 
-2\pi\gamma{\bf e}_z\int_{z_{end}}^{z} \left[h/(1+h_z^2)^{1/2}\right]_z dz
= -2\pi\gamma{\bf e}_z h/(1+h_z^2)^{1/2},
\label{gamma}
\end{equation}
since the height $h(z,t)$ goes to zero at the end of the drop.
Thus we arrive at the exact relation 
\begin{equation}
\int_0^{h(z,t)}{\bf e}_z\sigma r dr = 
-\gamma{\bf e}_zh/(1+h_z^2)^{1/2}
\label{t1}
\end{equation}
for the total force on the cross section of the drop.

Using the definition of the stress tensor
(\cite{LL}), the ${\bf e}_z$-component of (\ref{t1}) can be rewritten as 
\begin{equation}
\int_0^{h(z)}p(r,z) r dr + 2\eta v_r(h(z),z) = 
\gamma h/(1+h_z^2)^{1/2}, 
\label{t2}
\end{equation}
where $p(r,z)$ is the pressure, $\eta$ the viscosity of the liquid,
and $v_r(r,z)$ the radial component of the velocity. 

We want to show that $v_r(h(z),z)$ is always positive at local minima $h_{min}$
of $h(z)$, which would mean that the minimum radius can only increase,
so breakup can never occur. We are not able to compute the integral over the 
pressure in the most general case, so we concentrate on the case that
$h_{min}$ is small, as it must be if breakup were to occur. In that case
the radial dependence of the pressure is well described by a constant, 
which is (\cite{E97})
$p(r,z) \approx \gamma\kappa + 2\eta v_r(h,z)/h$. Using this expression
for the pressure, (\ref{t2}) turns into
\begin{equation}
v_r(h(z),z) = 
(\gamma/6\eta)\left(1/(1+h_z^2)^{1/2}+h h_{zz}/(1+h_z^2)^{3/2}\right).
\label{lub}
\end{equation}
At a local minimum of $h$ $h_{zz}$ is positive, 
making $v_r$ positive, so $h_{min}$
is {\it increasing} in time and breakup is impossible. 

We have been careful to invoke (\ref{lub}) only in the region around
$h_{min}$, where it is sure to hold. In fact, (\ref{lub}) is equivalent
to the one-dimensional lubrication-type description, which was found
to often work surprisingly well throughout a fluid drop or filament
(\cite{E97,B02}). Here, we find this observation confirmed, as illustrated
in figure \ref{fig1} by superimposing the lubrication calculation
(dashed lines) onto the full numerical calculation. No adjustable 
parameter was introduced in the comparison. It follows from
(\ref{lub}) and is illustrated by figure \ref{fig1}, that the minimum 
radius as given by the lubrication approximation must be monotonically 
increasing. We suspect that the same holds true for the full Stokes 
equations, but at present we can't exclude a decreasing $h_{min}$
in cases where $h_{min}$ is not small. It would be interesting 
to extend our results to the case that an exterior fluid is present. 
Results presented by \cite{PZGS98} suggest that breakup does not occur
as long as the exterior viscosity is sufficiently small.

\begin{acknowledgments}
We gratefully acknowledge funding by the EPSRC.
\end{acknowledgments}

\end{document}